\documentclass[12pt]{iopart}
\usepackage{graphicx}
\usepackage{epsfig}
\usepackage{bm}
\usepackage{color}
\usepackage{float}
\usepackage{dcolumn}
\begin{document}
\input epsf

\title{Addressing spectroscopic quality of covariant density
       functional theory}

\author{A.\ V.\ Afanasjev  
\footnote{The correspondence should be addressed to afansjev@erc.msstate.edu}
}

\address{Department of Physics and Astronomy, Mississippi State 
University, Mississippi State, Mississippi 39762, USA}

\begin{abstract}
The spectroscopic quality of  covariant density functional 
theory has been accessed by analyzing the accuracy and theoretical 
uncertainties in the description of spectroscopic observables. Such 
analysis is first presented for the energies of the single-particle 
states in spherical and deformed nuclei. It is also shown that the 
inclusion of particle-vibration coupling improves the description 
of the energies of predominantly single-particle states
in medium and heavy-mass spherical nuclei. However, the remaining 
differences between theory and experiment clearly indicate missing 
physics and missing terms in covariant energy density functionals. 
The uncertainties in the predictions of the position of 
two-neutron drip line sensitively depend on the uncertainties in
the prediction of the energies of the single-particle states.
On the other hand, many spectroscopic observables in well deformed 
nuclei at ground state and finite spin only weakly depend on the 
choice of covariant energy density functional.
\end{abstract}


% \submitto{\JPG}

%\maketitle

\section{Introduction}

  Among nuclear density functional theories (DFT), covariant density 
functional theory (CDFT) is one of most attractive since {\it covariant 
energy density functionals} (CEDF)  exploit basic properties of QCD 
at low energies, such as symmetries and the separation of scales 
\cite{LNP.641}. They provide a consistent treatment of the spin degrees 
of freedom, they include the complicated interplay between the large 
Lorentz scalar and vector self-energies induced on the QCD level by 
the in-medium changes of the scalar and vector quark condensates 
\cite{CFG.92,KW.08}. Their local scalar and vector fields appear in the 
role of local relativistic Kohn-Sham potentials \cite{SW.97,FS.00}. 
This class of successful scalar-vector mean field models can also be 
considered as a phenomenological realization of chiral effective field 
theories \cite{FKVW.06,KW.10}. The CEDF is approximated with powers 
and gradients of auxiliary meson fields or nucleon densities. The 
building of the covariant energy density functional in the context 
of effective field theory allows error estimates to be made, provides 
a power counting scheme which separates long- and short-distance 
dynamics, and therefore, removes model dependences from self-consistent 
mean field approach \cite{FS.00-1}.

  In addition, relativistic treatment of the DFT problems offers substantial 
advantages over non-relativistic approach. First of all, the spin-orbit 
interaction emerges in a natural way in the CDFT 
\cite{SW.86,Ring1996_PPNP37-193} and the results of the calculations 
are close to experiment \cite{LA.11,BRRMG.99}. On the contrary, 
spin-orbit interaction is always parametrized in non-relativistic DFT 
\cite{BHP.03}.
% and the calculated spin-orbit splittings deviate more from 
%the experiment than those obtained in the CDFT \cite{BRRMG.99}. 
Second, covariant energy density functionals include  {\it nuclear magnetism} 
\cite{KR.89}, i.e. a consistent description of currents and time-odd mean 
fields important for odd-mass nuclei \cite{AA.10}, the excitations with 
unsaturated spins, magnetic moments \cite{HR.88} and nuclear rotations 
\cite{AR.00,TO-rot}. Because of Lorentz invariance no new adjustable 
parameters are required for the time-odd parts of the mean fields in 
CDFT. The effects of nuclear magnetism are most pronounced in
rotating nuclei \cite{AR.00,TO-rot}. The fact that the 
properties of rotating nuclei are well described in the CDFT
calculations (see Refs.\ \cite{VALR.05,AO.13} and references
therein and Sect.\  \ref{ROT} in this paper) strongly suggests
that the effects of nuclear magnetism are correctly (as compared with 
experiment) reproduced. Moreover, they show only weak dependence on 
CEDF  \cite{AA.10,TO-rot}. These are important features which decrease 
model dependence of some spectroscopic observables (such as the moments of 
inertia). In contrast, several prescriptions (native, gauge and Landau 
\cite{SDMMNSS.10}) exist for the description of time-odd mean fields in 
non-relativistic Skyrme DFT. As a consequence, their impact on physical 
observables such as binding energies of one-quasiparticle configurations 
of odd-mass nuclei \cite{SDMMNSS.10} and the moments of inertia \cite{DD.95} 
is not uniquely defined.  

  Of course, at present, all attempts to derive CEDF's directly 
from the bare forces~\cite{BT.92,HKL.01,SOA.05,HSR.07} do not reach 
the required accuracy; the same is true also for non-relativistic 
%EDF's \cite{UNEDF2}. 
EDF's \cite{SKBDFGS.10}. Even for most microscopically based CEDF 
DD-ME$\delta$ \cite{DD-MEdelta}, four parameters are fitted to 
finite nuclei.
Considering phenomenological content of modern EDF's, it is 
important to estimate theoretical uncertainties in the EDF 
parameters and in the description of physical observables. 
It was suggested in Refs.\ \cite{RN.10,DNR.14} 
to use the methods of information theory for that purpose. These 
uncertainties come from the selection of the form of EDF as well 
as from the fitting protocol details, such as the selection of the 
nuclei under investigation, the physical observables, or the 
corresponding weights. Some of them are called {\it statistical 
errors} and can be calculated from a statistical analysis during 
the fit, others are {\it systematic errors}, such as for instance 
the form of the EDF under investigation, which are much more 
difficult to estimate because of possible missing physics \cite{DNR.14}. 
This is especially true because of the current bias on the use of 
bulk properties (masses, radii, neutron skins) in the fit of the 
non-relativistic and covariant EDF's \cite{AARR.14}. For example, 
spectroscopic (single-particle) information is never used in the 
fit of CEDF's. On the contrary, the limited information on the 
splitting of spin-orbit doublets is always employed in the fit of 
non-relativistic EDF's.

  The current paper aims on a review of spectroscopic quality of 
CDFT. In that respect it is useful to recall the definition of 
``nuclear spectroscopy''. There are several definitions available, 
and the one quoted below is a good representative example.
According to Ref.\ \cite{NS-def}, 
``{\it \bf Nuclear Spectroscopy} {\it is a branch of nuclear physics that is 
concerned with the study of the discrete spectrum of nuclear states, 
namely, with the determination of energy, spin, parity, isotopic spin, 
and other quantum characteristics of the nucleus in the ground and 
the excited states.''}
Experimentally, these quantities are obtained by measuring either 
the $\gamma$-transitions between the states in the same nucleus or 
different decays ($\alpha$, $\beta^+$, $\beta^-$ etc)  between the 
states in different nuclei. Thus, in the opinion of the author, 
this definition has to be extended by adding ``{\it and the 
transitions and decays between the states}'' at its end.

  This definition of ``nuclear spectroscopy'' is presented here 
because much narrower definition, focused only on the accuracy 
of the description of the single-particle states in odd-mass nuclei, 
is frequently used in theoretical community when spectroscopic 
quality of specific theory is discussed. This narrow definition  
ignores collective excitations such as rotations and surface vibrations 
leading to rotational and vibrational bands; they are also studied by 
nuclear spectroscopy. In the present paper, the author will use the 
broad definition of ``nuclear spectroscopy'' when discussing 
spectroscopic quality of CDFT.

 The analysis of theoretical uncertainties relies on statistical 
methods. The application of such methods is more complicated for
spectroscopic physical observables than for  ground state observables 
due to a number of reasons. First, such observables (for example, 
the moment of inertia which describes the evolution of rotational 
band  and the $\gamma$-transitions between the $I$ and $I+2$ members 
of the band with spin) depend on external parameter (rotational 
frequency) so the calculations have to performed for a set of the 
values of this parameter. In addition, they have to be calculated 
in  three-dimensional computer codes which are numerically 
time-consuming. Second, time-odd mean fields have to be taken 
into account for the calculations of the energies of 
one-(quasi)particle states in odd-mass spherical \cite{LA.11} 
and deformed \cite{AA.10,AS.11} nuclei and the moments of 
inertia \cite{KR.93,AR.00,TO-rot} of nuclear configurations. Third, 
the blocking procedure has to be employed in odd-mass nuclei. 
However, it is frequently numerically unstable (see Ref.\ 
\cite{AS.11} and the discussion in Sect.\ V of Ref.\ 
\cite{AO.13}).

  It is very difficult to perform the analysis of {\it statistical 
errors} on a global scale since the properties of transitional and 
deformed nuclei have to be calculated repeatedly for different 
variations of original EDF. Thus, such statistical analysis has been 
performed mostly for spherical nuclei \cite{RN.10,KENBGO.13} or selected 
isotopic chains of deformed nuclei \cite{Eet.12}. In Ref.\ 
\cite{AARR.14}, a global analysis of theoretical uncertainties 
for the ground state observables of even-even nuclei has been 
performed in the RHB calculations with four state-of-the-art 
CEDF's. However, these uncertainties are only crude approximation
to the {\it systematic theoretical errors} discussed in
Ref.\ \cite{DNR.14} because of (i) limited choice of CEDF's and (b) 
possible similar missing physics in these functionals. It is clear 
that similar global analysis is not possible for spectroscopic 
observables because of the complexity of their calculations. 
However, as will be shown below the estimate of theoretical 
uncertainties and important physical conclusions related to
spectroscopic physical observables can be obtained employing 
smaller set of data. 

   The paper is organized as follows. The accuracy of the 
description of predominantly single-particle states in spherical 
nuclei will be discussed in Sec.\ \ref{SP-spher}. In Sect.\ 
\ref{SP-def}, I will analyse theoretical uncertainties in the 
description of deformed single-(quasi)particle states. The 
impact of the imperfections in the single-particle structure on 
the spectroscopic observables in rotating nuclei will be 
considered in Sect.\ \ref{ROT}. Sect.\ \ref{DRIP} is dedicated
to the analysis of how the uncertainties in the description of 
the single-particle structure affect the predictions for the 
position of two-neutron drip line. Finally, Sect.\ \ref{Concl} 
summarizes the results of this work.

%%%%%%%%%%%%%%%%%%%%%%%%%%%%%%%%%%%%%%%%%%%%%%%%%%%%%%%%%%%%%%
\section{Single-particle states in spherical nuclei}
\label{SP-spher}
%%%%%%%%%%%%%%%%%%%%%%%%%%%%%%%%%%%%%%%%%%%%%%%%%%%%%%%%%%%%%%

  The experimental data on predominantly single-particle states 
in odd-mass spherical nuclei neighbouring to doubly magic nuclei
has frequently been compared with model calculations on the mean
field level. Let me give few examples; the list is definitely not
complete. Such comparisons are performed in $^{208}$Pb for the NL1, 
NL3 and NLSH CEDF's in Fig.\ 20 of Ref.\ \cite{ALR.98}), for the 
NL3, NL-Z, NL-Z2, NL-VT1 CEDF's in Fig.\ 1 of Ref.\ \cite{BRRMG.99}, 
for the DD-PC1 CEDF in Fig.\ 18 in Ref.\ \cite{DD-PC1}, and for 
the PC-PK1, DD-PC1, PC-F1, PC-LA and NL3* CEDF's in Fig. 6 of Ref.\ 
\cite{PC-PK1}). For $^{132}$Sn such comparisons are presented for 
DD-PC1 in Fig. 18 of Ref.\ \cite{DD-PC1} and for the PC-PK1, 
DD-PC1, PC-F1, PC-LA and NL3* CEDF's in Fig. 6 of Ref.\ \cite{PC-PK1}. 
These results lead to the following observations:

\begin{itemize}

\item
the difference between the energies of some spherical subshells 
obtained with different CEDF's can come close to 2 MeV,

\item
the energies of some subshells may not be so much affected by 
the selection of CEDF,

\item 
as a consequence of these two observations, the relative 
energies of different spherical subshells strongly depend 
on the CEDF,

\item 
the existence of large shell gaps does not depend on CEDF, 
however, their size is CEDF dependent,

\item 
the existence of smaller shell gaps and their size are 
strongly CEDF dependent.

\end{itemize}
As illustrated in Fig.\ \ref{266Pb} below, such features
are also valid for nuclei at the neutron-drip line. 

 In general, it is not so difficult to perform the analysis 
of the statistical errors in the model predictions of the energies
of the single-particle states of spherical nuclei at the mean 
field level in the spirit of Ref.\ \cite{DNR.14} (see, for example,
Fig.\ \ref{data-spread} below). However, it is already clear 
that such an analysis will provide only limited guidance because of 
two reasons. First, the states in odd-mass nuclei are strongly 
affected by coupling with vibrations \cite{LA.11,Litvinova2006_PRC73-044328}. 
Second, there are systematic differences between different classes 
of the models which, for example, lead to systematic differences in 
model predictions for superheavy nuclei (see discussion below).

%%%%%%%%%%%%%%%%%%%%%%%%%%%%%%%%%%%%%%%%%%%%%%%%%%%%%%%%%%%%%%
\begin{figure*}[ht]
\includegraphics[width=16.0cm]{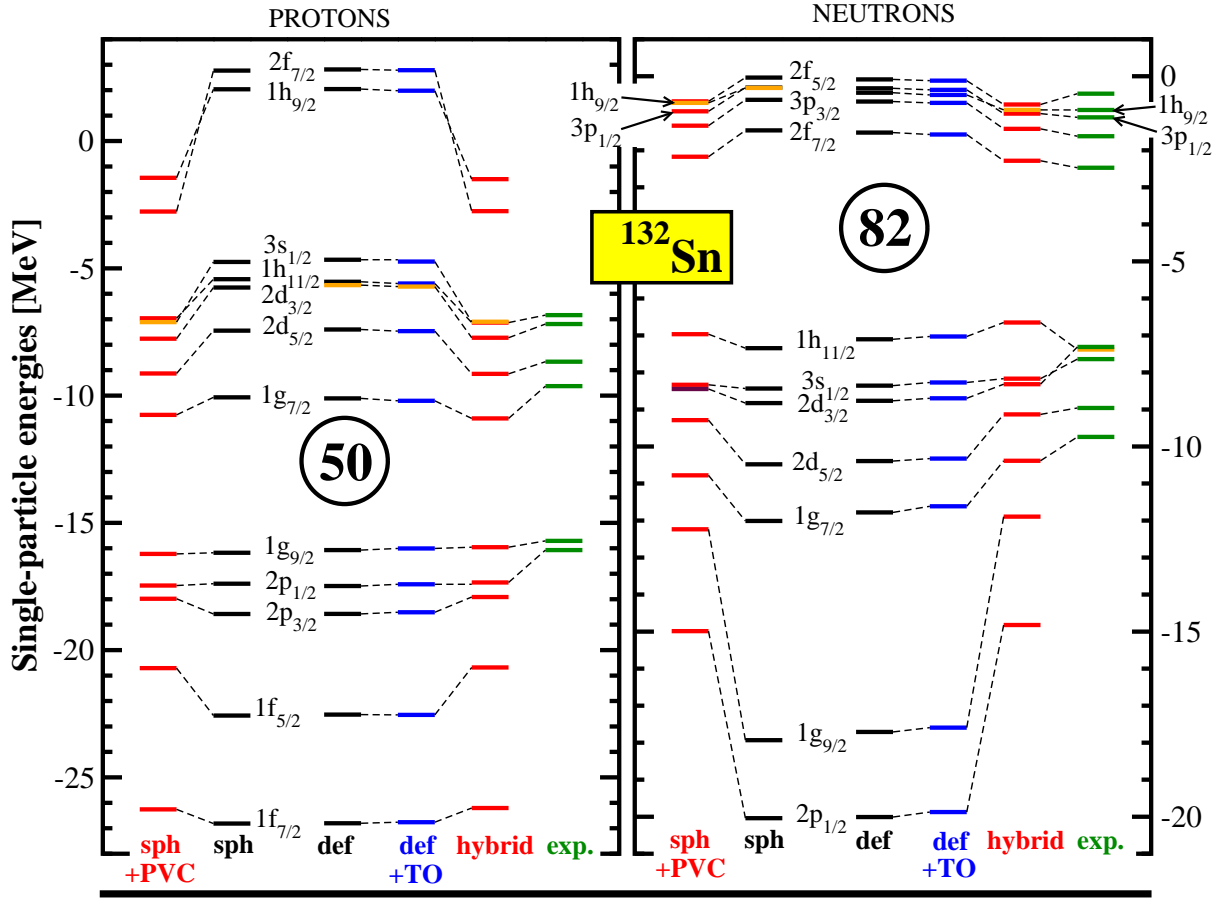}
\caption{Spectra of $^{132}$Sn and its
neighboring odd nuclei. Column 'sph' shows the 
single-particle spectra obtained in spherical CDFT 
calculations of $^{132}$Sn. Column 'sph+PVC' shows the 
spectra obtained in spherical calculations within the PVC 
model. Columns ``def'', ``def+TO'', ``hybrid'' and ``exp'' 
show one-nucleon separation energies \protect\cite{LA.11}. 
%defined according to Eqs.\ (\ref{Eq-part}) and
%(\ref{Eq-hole}). 
Column ``def'' is based on the results of triaxial
CRMF calculations with no time-odd mean fields. These 
fields are included in the calculations the results 
of which are shown in column ``def+TO''. The corrections 
due to PVC are added in column ``hybrid''. 
In order to distinguish 
overlapping levels, orange and then maroon colors are used for the levels 
in addition to their standard color used in a given column. From Ref.\ 
\cite{LA.11}.}
\label{132Sn-spectra}
\end{figure*}
%%%%%%%%%%%%%%%%%%%%%%%%%%%%%%%%%%%%%%%%%%%%%%%%%%%%%%%%%%%%%%

 Fig.\ \ref{132Sn-spectra} shows that {\it on average} the 
inclusion of particle-vibration coupling (PVC) substantially 
improves the description of the spectra in $^{132}$Sn.
A systematic analysis of Ref.\ \cite{LA.11} performed within
the relativistic PVC model with the NL3* CEDF \cite{NL3*} covers 
$^{56}$Ni, $^{132}$Sn and $^{208}$Pb. For these nuclei, average 
deviations per state $\Delta\varepsilon$ between calculated and 
experimental energies of the predominantly single-particle states 
are shown in Table \ref{Table-dev}. They are defined as 
\begin{equation}
\Delta \varepsilon = \frac{\sum_{i=1}^N 
|\varepsilon_i^{th}-\varepsilon_i^{exp}|}{N}
\end{equation}
where $N$ is the number of the states with known experimental 
single-particle energies, and $\varepsilon_i^{th}$ 
($\varepsilon_i^{exp}$) are calculated (experimental) energies 
of the predominantly single-particle states. One can see that 
the inclusion of PVC substantially improves the description of 
the single-particle states in $^{132}$Sn and $^{208}$Pb. The
same result has also been obtained in the $^{208}$Pb calculations 
with the NL3 CEDF \cite{Litvinova2006_PRC73-044328}.
On the contrary, PVC introduces no (small) improvement in the 
description of the proton (neutron) single-particle states 
of $^{56}$Ni. 

%%%%%%%%%%%%%%%%%%%%%%%%%%%%%%%%%%%%%%%%%%%%%%%%%%%%%%%%%%%%%%
\begin{table}[h]
\caption{Average deviations per state $\Delta\varepsilon$ 
between calculated and experimental energies of the single-particle 
states for a proton (neutron) subsystem of a given nucleus. The 
results obtained in the ``def+TO'' and ``hybrid'' calculational 
schemes  are shown (see caption of Fig.\ \ref{132Sn-spectra} and 
Ref.\ \cite{LA.11} for details).}
\begin{center}
\begin{tabular}{|c|c|c|} \hline
Nucleus/subsystem  &  $\Delta\varepsilon_{def+TO}$ [MeV] & $\Delta\varepsilon_{hybrid}$ [MeV] \\ 
\hline
 $^{56}$Ni/proton    &   0.76    &  0.77    \\
 $^{56}$Ni/neutron   &   0.89    &  0.71   \\
 $^{132}$Sn/proton   &   1.02    &  0.68   \\
 $^{132}$Sn/neutron  &   0.89    &  0.39   \\
 $^{208}$Pb/proton   &   1.53    &  0.84    \\
 $^{208}$Pb/neutron  &   1.00    &  0.47    \\ \hline
\end{tabular}
\end{center}
\label{Table-dev}
\end{table}
%%%%%%%%%%%%%%%%%%%%%%%%%%%%%%%%%%%%%%%%%%%%%%%%%%%%%%%%%%%%%%%%%%%

  One can ask a question what is a reason for a such different 
behavior of PVC in light and medium/heavy nuclei. Note that 
similar behavior is also observed in Skyrme PVC calculations in 
which the PVC does (not) improve the description of predominantly 
single-particle states in medium/heavy (light) nuclei \cite{TDTC.14}. 
One of possibilities is related to the fact that contrary to medium 
($^{132}$Sn)/heavy($^{208}$Pb) nuclei the lighter nuclei are 
characterized by soft potential energy surfaces. 
%(Fig.\ \ref{pes}). 
For such nuclei, the description of collective phonons within random 
phase approximation may be not adequate and more sophisticated methods 
such as generator coordinate method may be required \cite{RS.80}.

%%%%%%%%%%%%%%%%%%%%%%%%%%%%%%%%%%%%%%%%%%%%%%%%%%%%%%%%%%%%%%
%\begin{figure}[ht]
%\begin{center}
%\includegraphics[width=10.0cm]{pes.eps}
%\caption{\bf Potential energy curves for indicated nuclei obtained
%in the axial RHB calculations with the NL3* CEDF. The total
%energies are normalized to zero at $\beta_2=0$. 
%}
%\end{center}
%\label{pes}
%\end{figure}
%%%%%%%%%%%%%%%%%%%%%%%%%%%%%%%%%%%%%%%%%%%%%%%%%%%%%%%%%%%%%%

This discussion clearly shows that single-particle 
observables are more complicated than bulk ones (such as masses, 
radii, neutron skin thicknesses etc) for which statistical error 
analysis is relatively simple \cite{RN.10,DNR.14}. 
A necessary condition for an  analysis of systematic errors is 
statistical independence of EDF's under consideration \cite{DNR.14}.
However, this condition is not satisfied in modern DFT's. The 
dominance of bulk observables and the ignorance of single-particle 
observables in the fitting protocols of EDF's leads to the bias 
towards former observables and possible missing physics and terms 
of EDF's. Different model assumptions also contribute to that.

The region of superheavy nuclei is a clear example where model 
biases (missing terms of EDF's and missing physics) should be 
analysed in detail and addressed before the analysis of statistical
errors is undertaken. This is illustrated by the fact that
the centers of the island of stability are located at different
shell gaps in different models. For example, it is located at $Z=120$ 
and $N=172$ in most of CEDF's \cite{BRRMG.99}. Although the $N=172$ 
gap is preferred, 
neutron gap at $N=184$ cannot be excluded \cite{A250,LG.14}. Similar 
conclusion has also been reached in the relativistic PVC calculations 
of Refs.\ \cite{LA.11,L.12}.  On the other hand, non-relativistic Skyrme 
DFT favors $Z=126$ and $N=184$ and macroscopic+microscopic method favors
$Z=114$ and $N=184$ \cite{BRRMG.99}. 
 
%%%%%%%%%%%%%%%%%%%%%%%%%%%%%%%%%%%%%%%%%%%%%%%%%%%%%%%%%%%%%%
\section{Single-particle states in deformed nuclei}
\label{SP-def}
%%%%%%%%%%%%%%%%%%%%%%%%%%%%%%%%%%%%%%%%%%%%%%%%%%%%%%%%%%%%%%

 An essential difference between the phenomenological models 
based on the Woods-Saxon or Nilsson potentials and self-consistent 
DFT calculations is the fact that the phenomenological potentials 
are fitted to experimental single-particle energies. As a consequence, 
they well describe the single-particle spectra in deformed systems. 
On the contrary, no single-particle information is used in the fit 
of CEDF's. In the non-relativistic EDF's, the strength of the 
spin-orbit force is typically fitted to experimental data on 
spin-orbit splittings.

%%%%%%%%%%%%%%%%%%%%%%%%%%%%%%%%%%%%%%%%%%%%%%%%%%%%%%%%%%%%%%%%%%%%%%%%%%
\begin{figure}
\centerline{\psfig{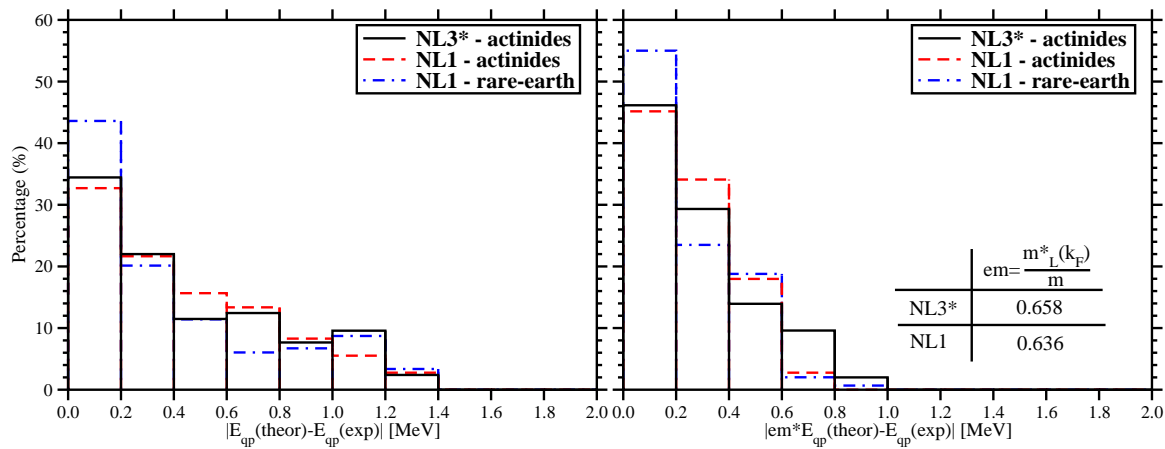}}
%\vspace*{8pt}
\vspace{-0.3cm}
\caption{(left panel) The distribution of the deviations of the calculated
energies $E_{qp}(theor)$ of one-quasiparticle states from experimental
ones $E_{qp}(exp)$. The vertical axis shows the percentage of the states
which deviate from experiment by the energy deviation range (the width of
bar) specified on horizontal axis. (right panel) The same as in left panel,
but for the case when the energy scale of theoretical spectra is corrected
for low Lorentz effective mass. Based on Figs.\ 2  and 3  of Ref.\ 
\protect\cite{AS.11}.}
\label{Dev-stat}
\end{figure}
%%%%%%%%%%%%%%%%%%%%%%%%%%%%%%%%%%%%%%%%%%%%%%%%%%%%%%%%%%%%%%%%%%%%%%%%

  Despite extensive use of the DFT's to the description of nuclear
phenomena, only recently few attempts to understand the accuracy
of the description of the single-particle spectra in deformed 
systems within the DFT framework have been undertaken. Restricted 
in scope investigations of experimental spectra in deformed odd 
nuclei have been performed in Skyrme \cite{SDMMNSS.10} and Gogny 
\cite{RSR.10} DFT. A statistical analysis of Ref.\ \cite{AS.11},
performed in the CDFT with NL1 and NL3* CEDF's, represents the 
most extensive attempt to understand the systematic errors in 
the description of deformed one-quasiparticle states.

%%%%%%%%%%%%%%%%%%%%%%%%%%%%%%%%%%%%%%%%%%%%%%%%%%%%%%%%%%%%%%%%%%%%%%%%%%
\begin{figure}
\centerline{\psfig{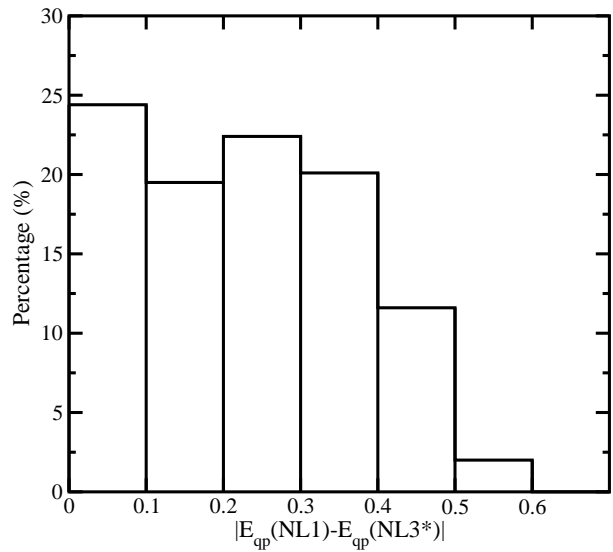}}
\vspace{-0.3cm}
\caption{The distribution of the differences between the energies 
$E_{qp}$ of one-quasiparticle states obtained in the RHB
calculations with the NL1 and NL3* CEDF's. The vertical axis shows the 
percentage of the states which fall into the energy difference 
range (the width of the bar) specified on horizontal axis.}
\label{Dev-stat}
\end{figure}
%%%%%%%%%%%%%%%%%%%%%%%%%%%%%%%%%%%%%%%%%%%%%%%%%%%%%%%%%%%%%%%%%%%%%%%%

  A statistical analysis \cite{AS.11} of the discrepancies between 
calculated and experimental energies of one-quasiparticle states 
in the ground state minimum is presented in the left panel of 
Fig.\ \ref{Dev-stat}. One can see that in the actinide region 
only approximately 33\% of one-quasiparticle states are described 
with an accuracy better than 200 keV, and approximately 22\% with
an accuracy between 200 and 400 keV in the NL3* and NL1 CEDF's. The 
percentage of the states for a given range of deviations gradually 
decreases with increasing deviation between experiment and calculations. 
However, for some states the deviation of the calculated energy from 
experiment exceeds 1 MeV and can be close to 1.4 MeV. Fig.\ 
\ref{Dev-stat} also shows that with the NL1 CEDF the 1-qp energies 
in odd-proton rare-earth nuclei are somewhat better described as 
compared with actinide region. Otherwise, the distribution histograms 
for the deviations are similar in both regions and for both 
parametrizations.

  The distribution of the differences between the energies 
$E_{qp}$ of one-quasiparticle states obtained in the RHB
calculations with the NL1 and NL3* CEDF's is presented
in Fig.\ \ref{Dev-stat}. One can see that substantial 
differences in the description of one-quasiparticle states
exist between employed CEDF's. The fact that these differences
are smaller than those presented in Fig.\ \ref{data-spread} below for 
spherical nuclei are due to two facts. First, only two CEDF's 
are used in Fig.\ \ref{Dev-stat}, while substantially larger set of ten 
CEDF's is used in Fig.\ \ref{data-spread}. Second, while one-quasiparticle 
energies defined with respect of the energy of the ground state 
in odd-mass nucleus are used in Fig.\ \ref{Dev-stat}, the 
absolute single-particle energies are used in the creation 
of Fig.\ \ref{data-spread}. 

  It is clear that the spectroscopic quality of the description 
of the single-particle spectra of the current generation of the DFT 
models (both relativistic and non-relativistic ones) is lower 
than the one achievable in the macroscopic+microscopic (MM) method.
In part, this is a consequence of different philosophies realized in 
the DFT and MM methods. It is well known that experimental 
``single-particle'' states are not mean-field states; their wave 
functions are fragmented and always contain the admixtures from 
vibrational phonons. In odd mass nuclei, the weights of these 
admixtures are generally low for ground states but increase with 
increasing excitation energy of the level relative to the ground state 
\cite{GISF.73,Sol-book}. By fitting 
the parameters  of phenomenological potentials to the energies of 
dominant single-particle states, the MM models effectively include 
vibrational corrections into these potentials but only on the
level of the energies and not on the level of the wavefunctions. 
As a consequence, these potentials are characterized by an effective 
mass of the nucleon at the Fermi level $m^*(k_F)/m\approx 1.0$ which 
reproduces experimental level density. However, for these
potentials the inclusion of the coupling to vibrations will lead 
to double counting of vibrational contribution in the energies and 
an effective mass of $m^*(k_F)/m\approx 1.4$ \cite{DDSMBB.00}.

  On the contrary, single-particle levels 
are not adjusted to experiment in DFT's since their 
functionals are fitted mainly to bulk and neutron matter 
properties. As a consequence, most of them, in particular 
Gogny and relativistic functionals, are characterized by 
low effective mass of the nucleon (the Lorentz mass for 
the case of CDFT \cite{JM.89}), and calculated single-particle 
states do not effectively include vibrational corrections. 
A low effective mass leads to a stretching of the theoretical 
single-particle energy scale as compared with experiment, 
and, thus, to larger deviations between theory and experiment 
for deformed one-quasiparticle states (left panel of Fig.\ 
\ref{Dev-stat}). To cure this problem one should go beyond 
the mean field approximation and supplement CDFT by 
particle-vibrational coupling (PVC). So far, this has been 
done only in spherical nuclei (see discussion in Sect.\ 
\ref{SP-spher}), for which it was shown that in the presence 
of PVC (i) calculated spectra of dominant single-particle 
states compress and come closer to experimental ones and (ii) 
effective mass of the nucleon comes closer to 1.

  A similar compression of calculated spectra is expected also
in deformed nuclei. However, so far, no PVC model based on the 
DFT framework has been developed for such nuclei. The analysis of 
Ref.\ \cite{AS.11} suggests that {\it on average} the expected 
compression of single-particle spectra can be achieved via a 
rescaling of one-quasiparticle (1-qp) energies by the Lorentz 
effective mass. The impact of such an energy rescaling on the 
distribution of the deviations between theory and experiment is 
shown in the right panel of Fig.\ \ref{Dev-stat}; more than 75\% 
of the states are described with an accuracy better than 400 keV.
This is a typical accuracy of the description of the energies of 
deformed 1-qp states within phenomenological potentials 
\cite{JSSJ.90,PS.04}. Although this energy rescaling is somewhat 
schematic and assumes that the effect of PVC is identical in 
spherical and deformed nuclei, it clearly illustrates that PVC, 
leading to an increase of the effective mass, could also improve 
the description of experimental spectra as compared with mean 
field results.

%%%%%%%%%%%%%%%%%%%%%%%%%%%%%%%%%%%%%%%%%%%%%%%%%%%%%%%%%%%%%%
\section{Rotational structures}
\label{ROT}
%%%%%%%%%%%%%%%%%%%%%%%%%%%%%%%%%%%%%%%%%%%%%%%%%%%%%%%%%%%%%%

 Considering existing inaccuracies in the description of 
the energies of the single-particle states, it is important 
to understand how they affect other physical observables of 
interest, especially the ones of collective nature in which 
many single-particle states contribute. It turns out that 
these inaccuracies do not affect appreciably many collective 
observables in the situations when potential energy surface 
as a function of collective variable is well developed. For 
example, the fission barriers in actinides are described 
accurately despite existing uncertainties in the description 
of the single-particle states and low effective mass of nucleon 
\cite{AAR.10,LZZ.12,PNV.12,AAR.12-ijmpe}. Moreover, the CDFT 
theory is the only DFT theory which provides such level of 
accuracy without a fit of the EDF parameters to the fission 
barriers or fission isomer energies \cite{AAR.13-epj}.

%%%%%%%%%%%%%%%%%%%%%%%%%%%%%%%%%%%%%%%%%%%%%%%%%%%%%%%%%%%%%%%%
\begin{figure*}[ht]
\begin{center}
\includegraphics[width=16.0cm,angle=0]{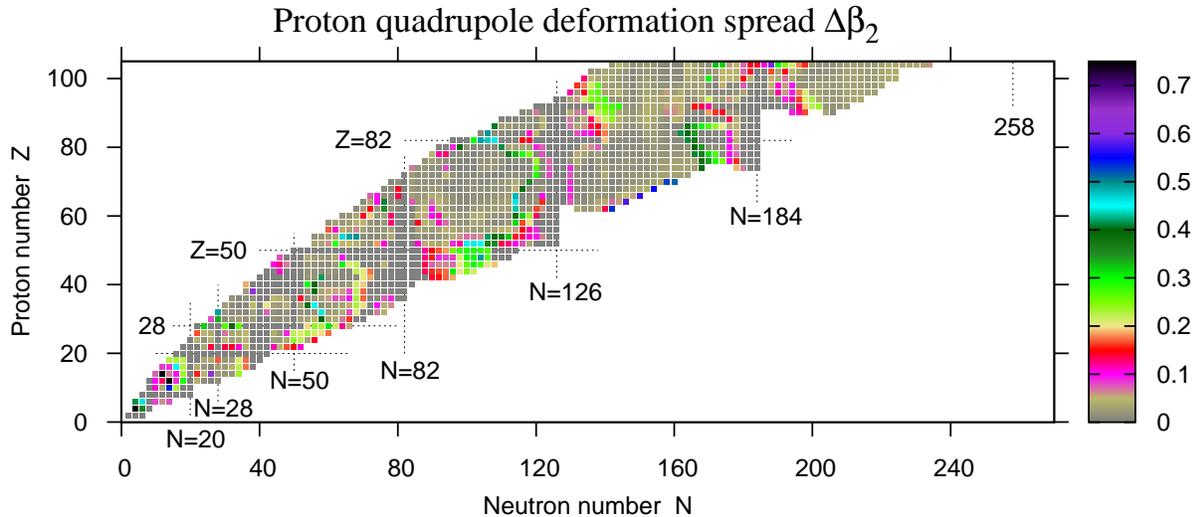}
\end{center}
\caption{Proton quadrupole deformation spreads
$\Delta \beta_2(Z,N)$ as a function of proton and neutron number.
$\Delta \beta_2(Z,N)=|\beta_2^{\rm max}(Z,N)-\beta_2^{\rm min}(Z,N)|$,
where $\beta_2^{\rm max}(Z,N)$ and $\beta_2^{\rm min}(Z,N)$ are the
largest and smallest proton quadrupole deformations obtained
with four employed CEDF's (NL3*, DD-ME2, DD-ME$\delta$ and DD-PC1)
for  the $(Z,N)$ nucleus at the ground state. From 
Ref.\ \protect\cite{AARR.14}.
See top panel of Fig.\ \protect\ref{chart-shade} for absolute 
values of the $\beta_2$-deformations calculated with DD-PC1.}
\label{Edif-charge-deform}
\end{figure*}
%%%%%%%%%%%%%%%%%%%%%%%%%%%%%%%%%%%%%%%%%%%%%%%%%%%%%%%%%%%%%%%%%

 Another process of interest which provides an important 
test of spectroscopic quality of the EDF is the rotation. The 
sequence of the states connected by stretched $E2$-transitions 
are formed in the case of the rotation of deformed electric 
charge distribution. The evolution of the energies of these 
transitions  and their strengths (in terms of B($E2$)) with 
spin  are frequently described in terms of the evolution 
of kinematic $J^{(1)}$ or/and dynamic $J^{(2)}$ moments
of inertia and charge ($Q_0$) or transition ($Q_t$) quadrupole 
moments with rotational frequency $\Omega_x$. So far the
systematic analysis of theoretical  uncertainties in the 
description of these spectroscopic observables has only 
been performed in actinides and light superheavy nuclei in 
Refs.\ \cite{AO.13,A.14} with the NL1 and NL3* CEDF's (see 
also discussion of ground state deformations in next paragraph).  
Some comparative results obtained with the NL1, NL3 and NLSH CEDF's 
are also available in the $A\sim 60$ ($^{58}$Cu, $^{60}$Zn, 
and $^{62}$Zn) \cite{A60}, $A\sim 150$ ($^{143}$Eu, $^{151}$Tb,
$^{151}$Dy and $^{152}$Dy) \cite{ALR.98} and $A\sim 190$ 
($^{194}$Pb and $^{194}$Hg) \cite{CRHB} regions of superdeformation.

%%%%%%%%%%%%%%%%%%%%%%%%%%%%%%%%%%%%%%%%%%%%%%%%%%%%%%%%%%%%%%%%%%
\begin{figure}[ht]
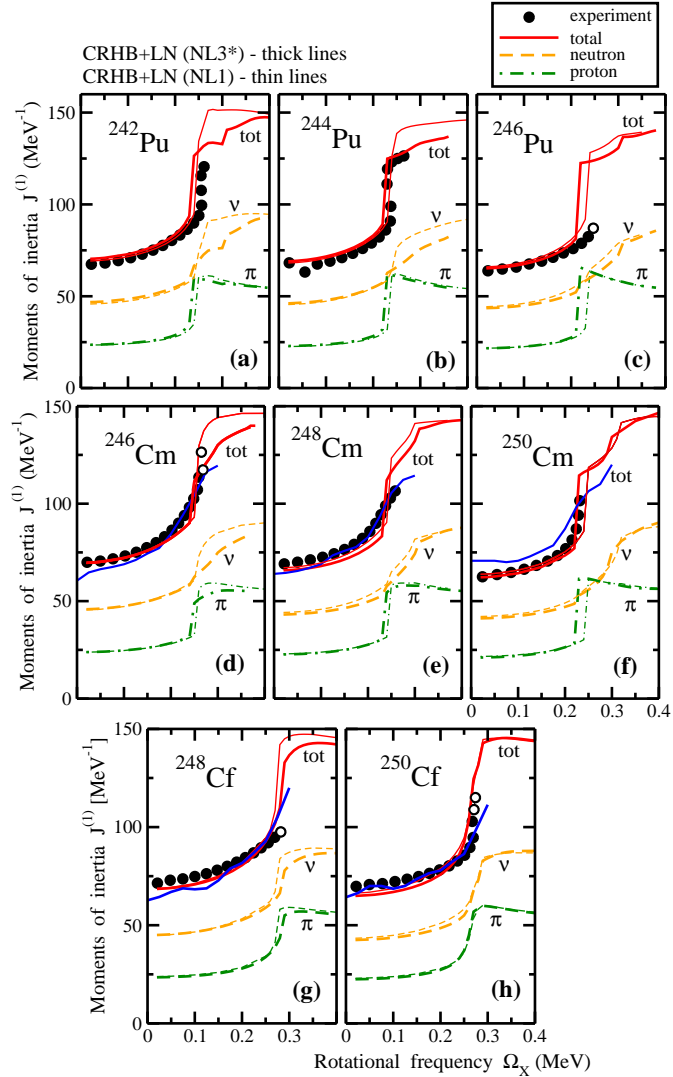

\begin{center}
\includegraphics[width=8.7cm]{Pu_CRHB_vs_CSM_mod}
\includegraphics[width=8.8cm]{Cm_back_CRHB_vs_CSM_mod}
\includegraphics[width=6.8cm]{Cf_CRHB_vs_CSM_mod}
\vspace{-0.4cm}
\caption{
The experimental and calculated kinematic moments of inertia 
$J^{(1)}$ of ground state rotational bands in indicated nuclei 
as a function of rotational frequency $\Omega_x$. Proton and 
neutron contributions to the kinematic moment of inerta are 
presented. Open circles are used for tentative experimental 
points. See Ref.\ \cite{A.14} for detailed comparison of 
these results.} 
\label{back_Pu}
\end{center}
\end{figure}
%%%%%%%%%%%%%%%%%%%%%%%%%%%%%%%%%%%%%%%%%%%%%%%%%%%%%%%%%%%%%%%%%%%

 As illustrated in Fig.\ \ref{Edif-charge-deform}, theoretical 
uncertainties in the prediction of ground state proton quadrupole 
deformations are rather small for the regions of well deformed 
nuclei such as rare-earth and actinides. Moreover, experimental 
data on $\beta_2$ in these regions are well (typically within the 
experimental uncertainties) described by CDFT (see Sect.\ IX in 
Ref.\ \cite{AARR.14}). Note that up to first backbending
the rotation does not appreciably change the equilibrium deformation 
in these regions (see, for example, Fig.\ 7 in Ref.\ \cite{A250}) 
and that this change only weakly depends on CEDF. Thus,
the results of Fig.\ \ref{Edif-charge-deform} strongly suggest 
that the deformations of ground state rotational bands up to
first backbending in these regions are well described in CDFT 
irrespective of employed CEDF. This similarity of calculated 
deformations exists also at higher spin, where again the differences 
between the $Q_t$ values obtained with different CEDF's for the
configuration of interest is typically within the experimental 
uncertainties (see also Refs.\ \cite{ALR.98,CRHB} for results
at superdeformation). Note that this result is strictly valid only 
for the configurations  which have 
well pronounced minima in potential energy surfaces. The existing 
systematics of the calculated transition quadrupole moments in 
rotating nuclei  \cite{VALR.05,A60,CRHB,A.12} shows that the CDFT 
well reproduces experimental data.

In addition to the deformation properties defining the strength 
of in-band $E2$-transitions, the spectroscopy of rotational bands 
shows up through the evolution of the moments of inertia with spin 
and band crossing features. Fig.\ \ref{back_Pu} illustrates both that 
up to the band crossing region the gradual rise of the kinematic moment 
of inertia $J^{(1)}$ is well reproduced in the CRHB+LN calculations and 
that the difference between two employed CEDF's (NL1 and NL3*) is rather
marginal. Systematic comparison of such results (compare Figs.\ 9 and 10 
in Ref.\ \cite{AO.13}) in actinides and light superheavy nuclei leads 
to the same conclusion. 
 
  The largest difference between the CEDF's shows up in the band 
crossing region. For example, the alignment of the $j_{15/2}$ neutrons 
in $^{242,244}$Pu and $^{246}$Cm proceeds in a gradual (sharp) way 
in the band crossing region in the CRHB+LN calculations with the NL3* 
(NL1) CEDF (Figs.\ \ref{back_Pu} a, b and d). As a consequence,
the alignment gain in the band crossing is also different in two 
CEDF's. However, for other nuclei shown in Fig.\ \ref{back_Pu} these 
differences are smaller; sharp upbend takes place at sligthly 
different frequencies and the differences in the alignment gain
are typically marginal. More examples of such differences between
the results obtained with the NL1 and NL3* CEDF's can be 
found in the systematic study of Ref.\ \cite{AO.13}. 
The strength of the interaction between the $g$ and $S$ bands and 
the crossing frequency depends sensitively on the relative position 
of aligning high-$j$ orbital with respect to the quasiparticle 
vacuum \cite{F.priv}. Whether the alignment in the band crossing 
region proceeds in a gradual (gradual increase of $J^{(1)}$) or sharp 
(sharp upbend in $J^{(1)}$) way depends on whether the interaction 
strength between the $g$ and $S$  bands is strong or weak.  Although 
the CRHB+LN calculations reproduce well the band crossings in 
$^{242,244}$Pu and $^{246,250}$Cm, they fail to reproduce the gradual 
alignment in $^{248}$Cm. It follows from the comparison between theory 
and experiment that the interaction strength between the $g$ and $S$  bands
shows variations with particle number which are not always reproduced 
in model calculations. The inaccuracies in the description of the 
single-particle states is one of possible reasons for this discrepancy 
between theory and experiment which decreases the predictive power 
of the models in the band crossing region.

%%%%%%%%%%%%%%%%%%%%%%%%%%%%%%%%%%%%%%%%%%%%%%%%%%%%%%%%%%%%%%%%%%%%%%%%
\begin{figure*}[ht]
\includegraphics[width=16.0cm,angle=0]{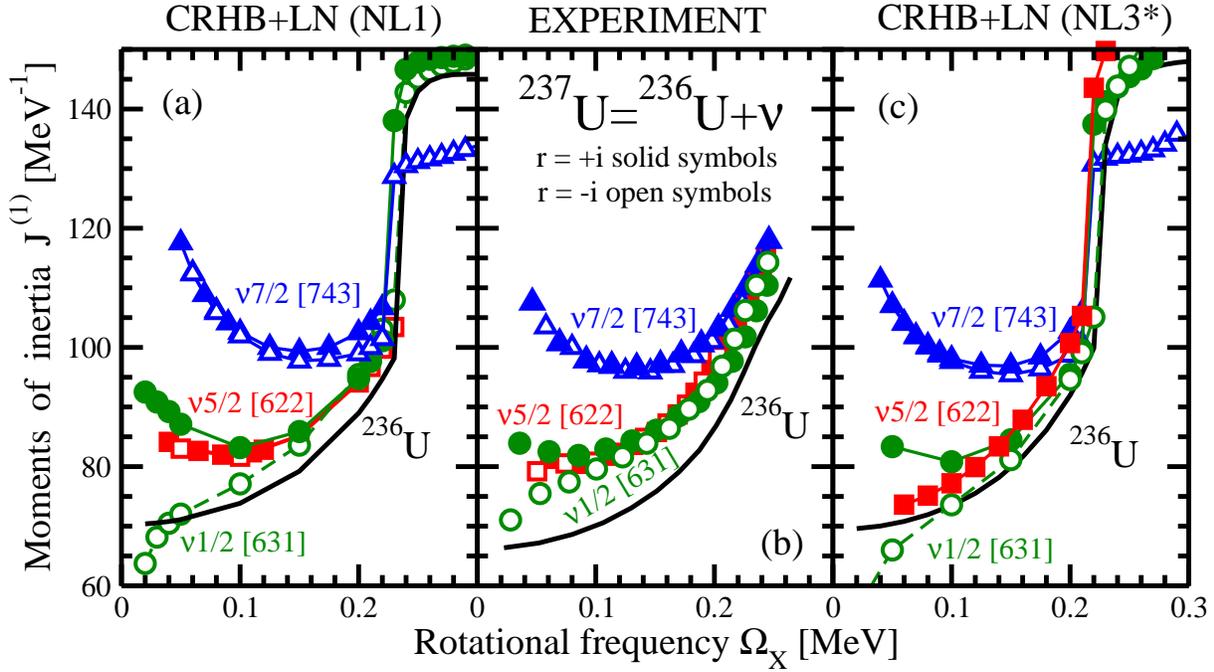}
\caption{Calculated and experimental kinematic moments of inertia 
$J^{(1)}$ of the indicated one-quasiproton configurations in the 
$^{237}$U nucleus and ground state rotational band in reference 
even-even $^{236}$U nucleus. Experimental data are shown in the 
middle panel, while the results of the CRHB+LN calculations with 
the NL1 and NL3* CEDF's in the left and right panels, 
respectively. The same symbols/lines are used for the same 
theoretical and experimental configurations. The symbols are used 
only for the configurations in odd-mass nucleus; the ground state 
rotational band in reference even-even nucleus is shown by solid 
black line. From Ref.\ \cite{AO.13}.} 
\label{J1-237U}
\end{figure*}
%%%%%%%%%%%%%%%%%%%%%%%%%%%%%%%%%%%%%%%%%%%%%%%%%%%%%%%%%%%%%%%%%%%%%%%%

  Rotational properties of one-quasiparticle configurations in odd-mass
nuclei provide an important information on the impact of odd particle/hole 
on alignment and pairing properties \cite{AO.13}. Considering that the 
energies of different deformed single-particle states with respect of the 
Fermi level and their relative energies depend on CEDF (see Sect.\ 
\ref{SP-def}), it is important to estimate theoretical uncertainties in the 
description of rotational properties of odd-mass nuclei emerging from the 
use of different CEDF's. Such a systematic estimate has been performed in 
Ref.\ \cite{AO.13}, and an illustrative example of $^{237}$U is shown in 
Fig.\ \ref{J1-237U}.

  For the $\nu 1/2[631]$ band in $^{237}$U, there is large separation 
between the $J^{(1)}$ values corresponding to the $(r=\pm i)$ branches at 
low frequency which gradually decreases and finally vanishes at high 
frequency (Fig.\ \ref{J1-237U}). This feature and the fact that the 
$(r=-i)$ branch has lower values of $J^{(1)}$ at low frequency are well 
reproduced in the calculations with the NL1 and NL3* CEDF's. However, the 
differences between the moments of inertia of these branches and the 
one in reference band of even-even nucleus $^{236}$U are underestimated.
The properties of two signature branches of the $\nu 7/2[743]$ rotational 
band such as the signature separation and its evolution  with frequency, 
their absolute $J^{(1)}$ values and evolution with frequency as well as 
their relative properties with respect of reference band in $^{236}$U are 
well reproduced in the calculations with both CEDF's.
The $\nu 5/2[622]$ rotational band is signature degenerate. This 
feature is well reproduced in the CRHB+LN(NL1) calculations. Only 
the $r=+i$ branch of this band has been obtained in the CRHB+LN(NL3*) 
calculations (Fig.\ \ref{J1-237U}c). However, the $\pi 5/2[523](r=\pm i)$ 
orbitals are signature degenerate in the frequency range of interest in 
the quasiparticle routhian diagram obtained with the NL3* CEDF. 
The absolute values of $J^{(1)}$ and their evolution with frequency are 
reproduced in model calculations. The NL1 CEDF somewhat better 
reproduces the properties of this band with respect of reference band in 
$^{236}$U than the NL3* CEDF which underestimates the increase 
of the $J^{(1)}$ values due to blocking of the $\nu 5/2[622](r=\pm i)$ 
orbitals.

  It is interesting to compare the CRHB+LN calculations \cite{AO.13} 
with the results of the cranked shell model in which the pairing 
correlations are treated by a particle-number conserving method (further 
CSM+PNC) \cite{ZHZZZ.12}. In the CSM+PNC model, the parameters of the Nilsson 
potential were carefully adjusted to the experimental energies of 
deformed one-quasiparticle states of actinides  and the experimental
deformations were used. Despite that the average accuracy of the 
description of rotational properties in the band crossing region 
and below is similar in both models (Fig.\ \ref{back_Pu}) \cite{A.14}.
However, an accurate description of the rotational properties in 
CRHB+LN \cite{AO.13} is achieved in a more consistent way than in 
the CSM-NPC  model. For example, contrary to CRHB+LN the accurate 
description of odd-mass nuclei in CSM+PNC model requires  a different 
pairing  strength as compared with even-even ones \cite{ZHZZZ.12}.

  Despite the fact that single-particle energies are described less 
accurately in CDFT as compared with the MM method, other aspects of the 
single-particle motion such as (i) deformation polarization effects 
induced by particle or hole (measured in terms 
of relative [or differential] transition quadrupole moments $\Delta Q_t$ 
of two bands \cite{ALR.98}) and (ii) alignment properties of 
single-particle orbital in rotating potential (measured by effective 
(relative) alignments \cite{Rag.93} of two compared bands)
are better described in CDFT. 
Indeed, the $\Delta Q_t$ values are well described in superdeformed 
rotational bands of the $A\sim 140-150$ mass region
in CDFT \cite{ALR.98,Dy151};  the average deviation from
experiment is around 20\%. Similar (but somewhat 
less accurate because of the role of pairing) results have been obtained 
also in the $A\sim 130$ mass region of high- and superdeformation 
\cite{L.02,MADLN.07}. The MM method based on 
the Nilsson potential describes deformation polarization effects 
also reasonably well. However, it suffers from the fact that these 
effects are not uniquely defined \cite{KR.98,Eet.00}. 
Effective alignments are also on average better reproduced in the CDFT 
than in the cranked Nilsson-Strutinsky version of the MM approach 
based on phenomenological Nilsson potential, see 
comparisons presented in Refs.\ \cite{AR.00,A60,AF.05}. 

%%%%%%%%%%%%%%%%%%%%%%%%%%%%%%%%%%%%%%%%%%%%%%%%%%%%%%%%%%%%%%
\section{Neutron drip lines}
\label{DRIP}
%%%%%%%%%%%%%%%%%%%%%%%%%%%%%%%%%%%%%%%%%%%%%%%%%%%%%%%%%%%%%%

%%%%%%%%%%%%%%%%%%%%%%%%%%%%%%%%%%%%%%%%%%%%%%%%%%%%%%%%%%%%%%
\begin{figure}[ht]
\includegraphics[angle=0,width=16.0cm]{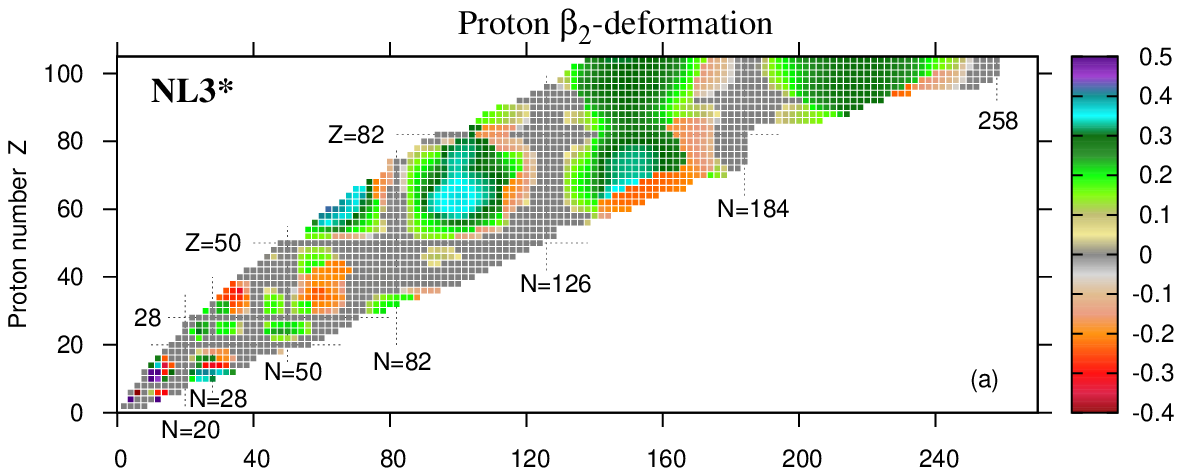}
\hspace{-0.5cm}
\includegraphics[angle=-90,width=16.0cm]{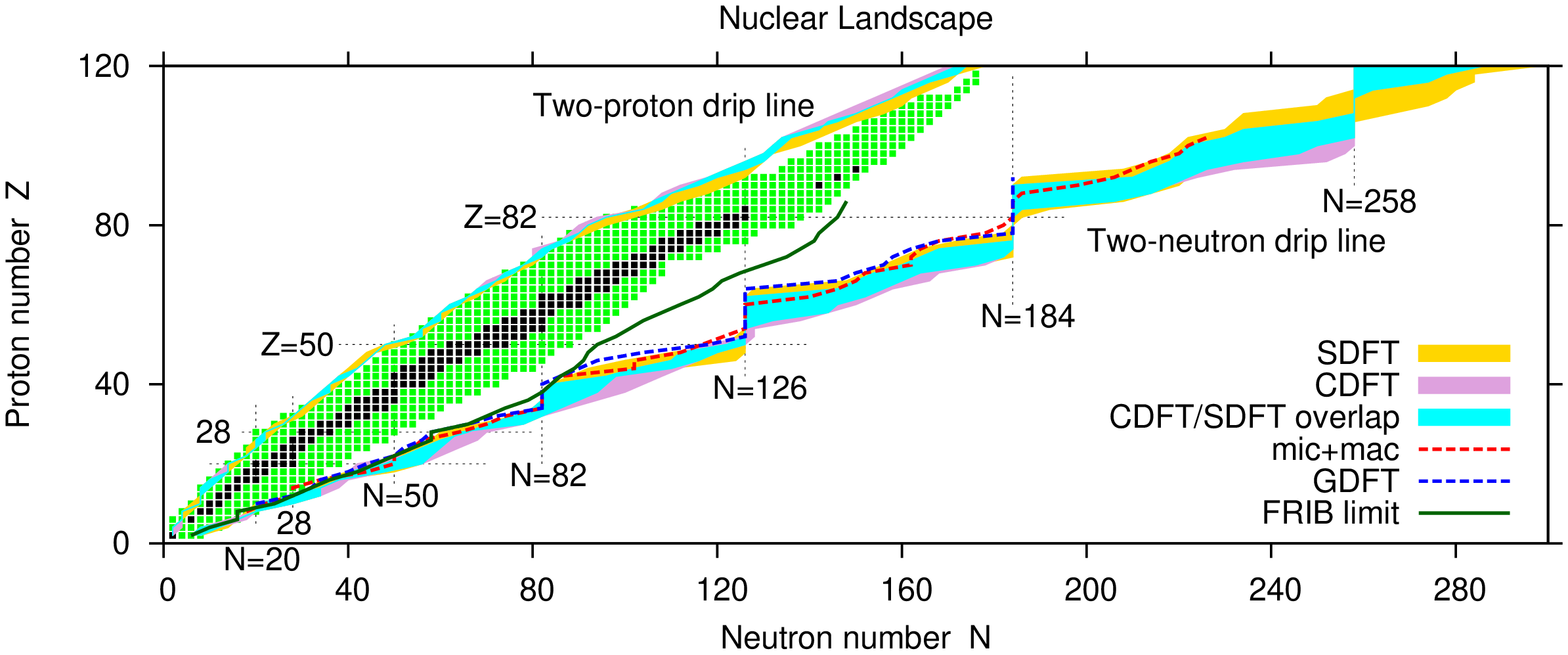}
\caption{(top panel) Charge quadrupole deformations $\beta_2$
obtained in the RHB calculations with the NL3* CEDF. From Ref.\ 
\cite{AARR.14}. (bottom panel)
Nuclear landscape as provided by state-of-the model 
calculations. The uncertainties in the definition of two-proton 
and two-neutron drip lines obtained in CDFT and Skyrme DFT (SDFT) 
are shown by shaded areas. They are defined by the extremes of the 
predictions of the corresponding drip lines obtained with different 
functionals. The blue shaded area shows the area where the CDFT and 
SDFT results overlap. Non-overlapping regions are shown by dark 
yellow and plum colors for SDFT and CDFT, respectively. 
The two-neutron drip lines obtained by microscopic+macroscopic (FRDM 
\protect\cite{MNMS.95}) and Gogny D1S DFT \protect\cite{DGLGHPPB.10} 
calculations  are shown by dashed red and blue lines, respectively. 
Experimentally known stable and radioactive (including proton emitters) 
nuclei  are shown by black and green squares, respectively. Green solid
line shows the limits of nuclear chart (defined as fission yield 
greater than $10^{-6}$) which may be achieved with dedicated existence 
measurements at FRIB \protect\cite{S-priv.14}. Based on Fig. 4 of Ref.\ 
\protect\cite{AARR.13}.}
\label{chart-shade}
\end{figure}
%%%%%%%%%%%%%%%%%%%%%%%%%%%%%%%%%%%%%%%%%%%%%%%%%%%%%%%%%%%%%%%%%%

  The analysis of theoretical uncertainties in the prediction 
of the position of the neutron and proton drip-lines has recently 
attracted great interest \cite{Eet.12,AARR.13,AARR.14} because of 
the possibility to estimate the number of nuclei which may exist 
in nature. Fig.\ \ref{chart-shade} (bottom panel) shows the 
nuclear landscape which emerges from such an analysis performed 
in the framework of state-of-the-art non-relativistic and 
relativistic DFT's.
 
  One can see that the largest uncertainties exist in the 
position of two-neutron drip line. Inevitably, the question 
about possible sources of these uncertainties emerges. For
example, they were related to existing uncertainties in the 
definition of isovector properties of the EDF's (Ref.\ 
\cite{Eet.12}). Indeed, the isovector properties of an EDF 
impact the depth of the nucleonic potential with respect to 
the continuum, and, thus, may affect the location of two-neutron 
drip line. However, inaccurate reproduction of the depth of 
the nucleonic potential exist in modern CEDF's also in known 
nuclei (see discussion in Sect. IVC of Ref.\ \cite{LA.11}). 
Thus, they alone cannot explain observed features. The observed 
differences in the prediction of the position of two-neutron 
drip line cannot also be explained by underlying nuclear matter 
properties of EDF's \cite{AARR.14}.

\begin{figure}[ht]
\begin{center}
\includegraphics[angle=0,width=16.0cm]{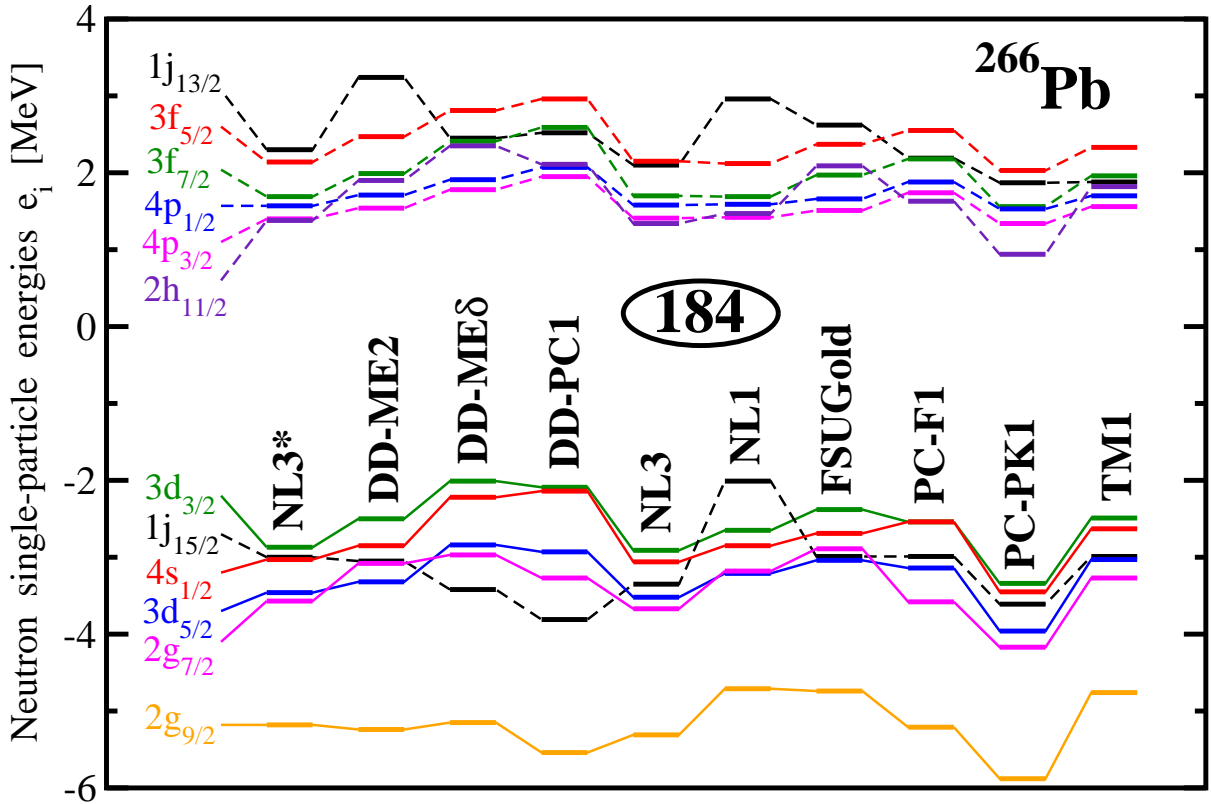}
\caption{Neutron single-particle states at spherical shape
in a $^{266}$Pb nucleus obtained with indicated CEDF's. Solid 
and dashed connecting lines are used for positive and negative 
parity states. Spherical gap at $N=184$ is indicated; all the 
states below this gap are occupied. For simplicity, only six
lowest states above and six highest states below the gap are 
shown.
}
\label{266Pb}
\end{center}
\end{figure}

 It was suggested in Ref.\ \cite{AARR.13} that the position of 
two-neutron drip line sensitively depends also on the underlying 
shell structure and the accuracy of the description of the 
energies of the single-particle states. Indeed, the shell structure 
effects are clearly visible in the fact that for some combinations 
of $Z$ and $N$ there is basically no (or very little) dependence 
of the predicted location of the two-neutron drip line on the 
EDF \cite{AARR.13,AARR.14} (see bottom panel of Fig.\ \ref{chart-shade}). 
Such a weak (or vanishing) dependence, seen in all model calculations, 
is especially pronounced at spherical neutron shell closures with 
$N=126$ and $184$ around the proton numbers $Z=54$ and $80$, 
respectively. In addition, a similar situation is seen in the 
CDFT calculations at $N=258$ and $Z\sim 110$. This fact is easy 
to understand because of the large neutron shell gap at the 
magic neutron numbers in all DFT's. This is illustrated in 
Fig.\ \ref{266Pb} where the magic $N=184$ shell gap has a 
significant size of around 4 MeV for all CEDF's. Note 
that only first four CEDF's of Fig.\ \ref{266Pb} were used in 
the definition of theoretical uncertainties in the position 
of two-neutron drip line in Fig.\ \ref{chart-shade}.   

\begin{figure}[ht]
\begin{center}
\includegraphics[angle=0,width=8.0cm]{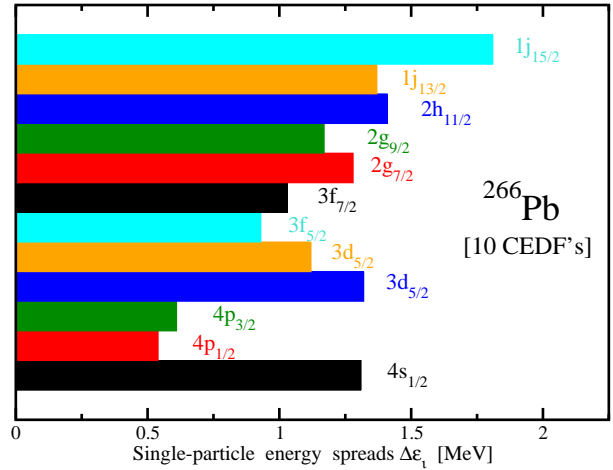}
\caption{The spreads $\Delta \epsilon_i$ for indicated neutron 
single-particle states in $^{266}$Pb. $\Delta \epsilon_i=
|\epsilon_i^{\rm max}-\epsilon_i^{\rm min}|$, where 
$\epsilon_i^{\rm max}$ and $\epsilon_i^{\rm min}$ are the
largest and smallest energies of a given single-particle 
state obtained with ten employed CEDF's for a given single-particle 
state.  The orbital angular momentum of the single-particle state 
increases on going from bottom to the top of the figure.  Based 
on the results presented in Fig.\ \ref{266Pb}.}
\label{data-spread}
\end{center}
\end{figure}

  Theoretical uncertainties in the description of the energies
of individual spherical orbitals shown in Fig.\ \ref{266Pb} 
are summarized in Fig.\ \ref{data-spread}. They are substantial 
and in the majority of the cases they exceed 1 MeV; these 
uncertainties are below 1 MeV only for three low-$j$ orbitals, 
namely, $4p_{1/2}$, $4p_{3/2}$ and $3f_{5/2}$. However, they are 
comparable with the ones observed in known nuclei (see the 
discussion in Sect.\ \ref{SP-spher}). These uncertainties 
definitely affect the position of two-neutron drip line and 
several factors discussed below play a role.

  First, the comparison of bottom and top panels of Fig.\ 
\ref{chart-shade} shows that there is a close correlation between 
the nuclear deformation at the neutron-drip line and the 
uncertainties in the prediction of this line. The regions of 
large uncertainties corresponds to transitional and deformed 
nuclei. Again this is caused by the underlying level densities 
of the single-particle states. The spherical nuclei under discussion 
are characterized by large shell gaps and a clustering of highly 
degenerate single-particle states around them. Deformation removes 
this high degeneracy of the single-particle states and leads to a more 
equal distribution of the single-particle states with energy.

  Second, the large density of the neutron single-particle states
close to the neutron continuum leads to a small slope of two-neutron
separation energies $S_{2n}$ as a function of neutron number in the 
vicinity of two-neutron drip line for medium and heavy mass nuclei
(see Fig.\ 12 in Ref.\ \cite{AARR.14}). As discussed in details in 
Sec.\ VIII of Ref.\ \cite{AARR.14} this translates into (i) much 
larger uncertainties in the definition of the position of two-neutron 
drip line as compared with two-proton drip line and (ii) stronger 
dependence (as compared with two-proton drip line) of the predictions 
for the position of the two-neutron drip line on the accuracy of 
the description of the energies of the single-particle states.

  Third, the position of the cluster of the states above the
zero energy, the ordering of the single-particle states in 
this cluster and, in particular, the relative positions of 
low-$j$ and high-$j$ spherical orbitals (see Fig.\ \ref{266Pb})
play an important role. This can be illustrated by the 
analysis of the Rn $(Z=86)$ isotope chains, two-neutron drip 
lines of which are located at $N=206$ in the NL3* CEDF and
at $N=184$ in the DD-ME2, DD-ME$\delta$ and DD-PC1 CEDF's
(Table IV in Ref.\ \cite{AARR.14}). The spectra shown in 
Fig.\ \ref{266Pb} will be used in this analysis; the increase 
of proton number from $Z=82$ (Pb) up to $Z=86$ (Rn) leads to 
a more or less constant shift down (by approximately 400 keV) 
of the single-particle spectra. The strong dependence of
the position of predicted neutron-drip line on the energies of 
the single-particle states is seen in the fact that the energies 
of the lowest positive energy states obtained with the NL3* 
and DD-ME2 CEDF's (two left columns in Fig.\ \ref{266Pb}),
which are expected to be active in the $N\geq 184$ nuclei, 
differ by only 160 keV. However, this difference alone cannot
explain the twenty two (22 !!!) neutron difference in the 
predicted position of two-neutron drip line. It is clear that 
the presence of low-lying high-degeneracy $2h_{11/2}$ orbital 
(which is also deformation-driving) at low energy in the NL3* 
CEDF is important for an extension of the nuclear landscape 
up to $N=206$. On the contrary, this orbital is located 
at higher energy in the DD- CEDF's (Fig.\ \ref{266Pb}). This
together with higher energies of other positive energy 
states leads to the termination of nuclear landscape at $N=184$
in these CEDF's.

%%%%%%%%%%%%%%%%%%%%%%%%%%%%%%%%%%%%%%%%%%%%%%%%%%%%%%%%%%%%%%
%%%%%%%%%%%%%%%%%%%%%%%%%%%%%%%%%%%%%%%%%%%%%%%%%%%%%%%%%%%%%%%%%%

%%%%%%%%%%%%%%%%%%%%%%%%%%%%%%%%%%%%%%%%%%%%%%%%%%%%%%%%%%%%%%
%%%%%%%%%%%%%%%%%%%%%%%%%%%%%%%%%%%%%%%%%%%%%%%%%%%%%%%%%%%%%%%%%%

%%%%%%%%%%%%%%%%%%%%%%%%%%%
\section{Conclusion}
\label{Concl}
%%%%%%%%%%%%%%%%%%%%%%%%%%%

   In the current manuscript, the question of spectroscopic 
quality of the description of physical phenomena in the 
framework of covariant density functional theory has been 
considered by analysing the accuracy and theoretical 
uncertainties in the description of different physical 
observables; these uncertainties are related to {\it 
systematic errors} of Ref.\ \cite{DNR.14}. The following 
conclusions have been obtained:

\begin{itemize}
\item
 The current generation of the covariant energy density 
functionals has been obtained in the fitting protocols
based on bulk and nuclear matter observables; no single-particle 
observables have been used in these protocols. Observed spreads 
in the predictions of the energies of the single-particle 
states in spherical and deformed nuclei when different CEDF's
are used clearly indicate that such fitting protocols do not
sufficiently constraint such observables. In medium and 
heavy-mass spherical nuclei relativistic particle-vibration 
coupling improves the description of the spectra. However, the 
remaining differences between theory and experiment clearly 
indicate missing physics and missing terms of CEDF's.

\item
  To improve the description of the single-particle states 
new types of the fitting protocols including experimental data 
on the single-particle states are needed. At the current level 
of the development of CDFT, two types of the protocols could 
become feasible in near future. First, the protocols focused 
on spherical nuclei but which in addition to standard observables
include the calculations of the spectra of few nuclei within the 
relativistic particle-vibration coupling model. Second, the 
protocols focused on deformed nuclei but which also require
that for a large enough set of normal deformed odd nuclei the 
structure of the ground state is reproduced as a function of 
proton and neutron numbers. Such protocols avoid the effect of 
low effective mass of nucleon at the Fermi level since only the 
lowest in energy one-quasiparticle state in each nucleus are 
used. Such states are only  weakly affected by 
quasiparticle-phonon coupling \cite{GISF.73}, and, thus, can be 
reasonably well treated at the DFT level. At present, the ground 
states of odd-mass  nuclei are correctly reproduced in the DFT 
calculations less frequently as compared with the MM ones 
\cite{AS.11,BQM.07} and the improvement in that direction is
highly desirable.

\item
 Considering existing theoretical uncertainties in the description 
of the energies of the single-particle states it is important to 
understand which physical observables and in which situations 
are well or poorly described or predicted. Because of these 
uncertainties the predictive and
descriptive power of the models decreases substantially (they can even 
give ``false positive'' or ``false negative'' signal on the existence 
of specific phenomenon) in the 
situations when the details of the potential energy surfaces 
(PES) sensitively depend on the underlying single-particle
structure. As seen in Fig.\ \ref{Edif-charge-deform} and 
discussed in details in Ref.\ \cite{AARR.14} this takes place 
at zero spin in transitional nuclei, the PES of which is soft, and in nuclei 
characterized by prolate-oblate shape coexistence. At higher
spin, this takes place, for example, in chiral rotational bands,
the PES of which are characterized by extremely shallow minima 
($\sim 50$ keV) \cite{AS.11}.

\item 
 The uncertainties in the energies of the 
single-particle states become less important in nuclear systems
with pronounced minima in the potential energy surfaces. These 
are, for example, well deformed nuclei in the rare-earth region 
and actinides or superdeformed structures across the nuclear 
chart. The rotational structures in the actinides were used
to illustrate this feature. The calculated physical observables
of actinides such as the moments of inertia and  deformations 
only weakly depend on the selection of CEDF. The sensitivity 
of the results to the selection of CEDF is more pronounced in 
the paired band crossing region. Some aspects of the single-particle 
motion such as deformation polarization effects induced by particle 
or hole and the alignment properties of single-particle orbitals 
are better and more consistently described in the CDFT as compared 
with the models based on phenomenological Nilsson potential despite 
the fact that this potential better describes the energies of the 
single-particle states.

\item
   The predicted position of two-neutron drip line in the 
majority of the isotope chains sensitively depends on the 
description of energies of the single-particle states in 
neutron-rich nuclei. It was illustrated in spherical nuclei 
that the uncertainties of such description are comparable in 
known and neutron-rich (near two-neutron drip line) nuclei. 
This fact strongly suggests that the improvement of the 
description of the energies of the single-particle states 
in known nuclei will reduce the uncertainties in the 
prediction of neutron-drip lines.

\end{itemize}

%%%%%%%%%%%%%%%%%%%%%%%%%%%
\section{Acknowledgements}
%%%%%%%%%%%%%%%%%%%%%%%%%%%

  This work has been supported by the U.S. Department of Energy 
under the grant DE-FG02-07ER41459 and partially by an allocation 
of advanced computing resources provided by the National Science 
Foundation. The computations were partially performed on Kraken 
at the National Institute for Computational Sciences 
(http://www.nics.tennessee.edu/). The help of S.\ Agbemava and 
D.\ Ray in preparation of Figs.\ \ref{Dev-stat} and 
\ref{chart-shade} and useful discussions with P.\ Ring are 
greatly appreciated.  I would also like to thank H.\ Schatz for 
providing FRIB rates.

\section*{References}
\bibliographystyle{iopart-num}
\bibliography{references}

\end{document}